\documentclass[abbrv,aps,prl,twocolumn,showpacs,preprintnumbers,amsmath,amssymb,
superscriptaddress,nofootinbib]{revtex4}

\usepackage{graphicx} 
\usepackage{dcolumn}
\usepackage{bm}
\usepackage{epstopdf}
\usepackage{color}
\usepackage{ulem}

\begin{document}
\title {Strong nonlinear terahertz response induced by Dirac surface states in Bi$_2$Se$_3$ Topological Insulator}

\author{Flavio~Giorgianni}
\affiliation {INFN and Dipartimento di Fisica, Universit\`a di Roma "La Sapienza",
Piazzale A. Moro 2, I-00185 Roma, Italy}

\author{Enrica~Chiadroni}
\affiliation {Laboratori Nazionali di Frascati - INFN
via Enrico Fermi, 40
00044 Frascati, Italy}

\author{Andrea~Rovere}
\affiliation {INFN and Dipartimento di Fisica, Universit\`a di Roma "La Sapienza",
Piazzale A. Moro 2, I-00185 Roma, Italy}

\author{Mariangela~Cestelli-Guidi}
\affiliation {Laboratori Nazionali di Frascati - INFN
via Enrico Fermi, 40
00044 Frascati, Italy}

\author{Andrea Perucchi}
\affiliation {INSTM Udr Trieste-ST and Sincrotrone Trieste, Area Science Park, I-34012
Trieste, Italy}

\author{Marco~Bellaveglia}
\affiliation {Laboratori Nazionali di Frascati - INFN
via Enrico Fermi, 40
00044 Frascati, Italy} 
\author{Michele~Castellano}
\affiliation {Laboratori Nazionali di Frascati - INFN
via Enrico Fermi, 40
00044 Frascati, Italy} 
\author{Domenico~Di Giovenale}
\affiliation {Laboratori Nazionali di Frascati - INFN
via Enrico Fermi, 40
00044 Frascati, Italy}
\author{Giampiero~Di Pirro}
\affiliation {Laboratori Nazionali di Frascati - INFN
via Enrico Fermi, 40
00044 Frascati, Italy}
\author{Massimo~Ferrario}
\affiliation {Laboratori Nazionali di Frascati - INFN
via Enrico Fermi, 40
00044 Frascati, Italy}  
\author{Riccardo~Pompili}
\affiliation {Laboratori Nazionali di Frascati - INFN
via Enrico Fermi, 40
00044 Frascati, Italy}  
\author{Cristina~Vaccarezza} 
\affiliation {Laboratori Nazionali di Frascati - INFN
via Enrico Fermi, 40
00044 Frascati, Italy} 
\author{Fabio~Villa}
\affiliation {Laboratori Nazionali di Frascati - INFN
via Enrico Fermi, 40
00044 Frascati, Italy}

\author{Alessandro~Cianchi}
\affiliation {INFN and Dipartimento di Fisica, Universit\`a di Roma "Tor Vergata",
viale della Ricerca Scientifica 1, I-00133 Roma, Italy}

\author{Andrea~Mostacci}
\affiliation {INFN and Dipartimento S.B.A.I., Universit\`a di Roma "La Sapienza",
Piazzale A. Moro 2, I-00185 Roma, Italy and via A. Scarpa 16, I-00161 Roma, Italy}

\author{Massimo~Petrarca}
\affiliation {INFN and Dipartimento S.B.A.I., Universit\`a di Roma "La Sapienza",
Piazzale A. Moro 2, I-00185 Roma, Italy and via A. Scarpa 16, I-00161 Roma, Italy}

\author{Matthew~Brahlek} 
\affiliation{Department of Physics and Astronomy Rutgers, The State University of New Jersey 136 Frelinghuysen Road Piscataway, NJ 08854-8019 USA}

\author{Nikesh~Koirala} 
\affiliation{Department of Physics and Astronomy Rutgers, The State University of New Jersey 136 Frelinghuysen Road Piscataway, NJ 08854-8019 USA}

\author{Sean~Oh}
\affiliation{Department of Physics and Astronomy Rutgers, The State University of New Jersey 136 Frelinghuysen Road Piscataway, NJ 08854-8019 USA} 

\author{Stefano~Lupi}
\affiliation {INFN and Dipartimento di Fisica, Universit\`a di Roma "La Sapienza",
Piazzale A. Moro 2, I-00185 Roma, Italy}

\date{\today}

\maketitle

Electrons with a linear energy/momentum dispersion are called massless Dirac electrons and represent the low-energy excitations in exotic materials like Graphene and Topological Insulators (TIs).
Dirac electrons are characterized by notable properties like a high mobility, a tunable density and, in TIs, a protection against backscattering through the spin-momentum looking mechanism. All those properties make Graphene and TIs appealling for plasmonics applications. However, Dirac electrons are expected to present also a strong nonlinear optical behavior. This should mirror in phenomena like electromagnetic induced transparency (EIT) and harmonic generation. Here, we demonstrate that in Bi$_2$Se$_3$ Topological Insulator, an EIT is achieved under the application of a strong terahertz (THz) electric field. 
This effect, \textcolor{black}{concomitant determined by harmonic generation and charge-mobility reduction}, is exclusively related to the presence of Dirac electron at the surface of Bi$_2$Se$_3$, and opens the road towards tunable THz nonlinear optical devices based on Topological Insulator materials.

\subsection{Introduction}

Nonlinear optical phenomena have a crucial importance in modern physics giving rise to fundamental applications like coherent control of excitations in condensed matter and harmonic generation and frequency conversion in optically active materials \cite{Boyd,cavalleri1,cavalleri2}.
In this case the use of materials whose electromagnetic response can be fully controlled by an applied radiation field, plays a fundamental role in ultrafast electromagnetic pulse generation and shaping \cite{femto2}. These nonlinear optical properties, earlier discovered and studied in the visible and near-infrared range of the electromagnetic spectrum, have been successively investigated at infrared frequencies \cite{IR}, and their extension towards the terahertz region (1 THz=33 cm$^{-1}$=300 $\mu$m=4 meV), a spectral range which has seen recently a tumultuous technological and scientific development \cite{Rapporto-THz}, is highly desirable. Terahertz research has been oriented either on investigating novel radiation sources based on frequency conversion, optical rectification \cite{Hebling, Photoantenne, Vicario} and relativistic electrons \cite{Carr, Chiadroni, Andrea}, and in studying the properties of plasmon-based systems whose optical properties like absorption, dispersion, and scattering can be engineered at THz frequencies \cite{Padilla}.  
As a matter of fact, the discovery of natural materials showing exotic nonlinear THz properties could set a new agenda in photonic and plasmonic applications of terahertz radiation.

One of the first observations of a nonlinear THz effect has been achieved in doped semiconductors by means of a THz electric field in the hundred of kV/cm range \cite{nsemi, nsemi2}. In GaAs, for instance, free-charge carriers introduced by doping or thermally excited in the conduction band can be accelerated by the THz electric field. When the momentum gain is sufficiently high, carriers are scattered from the bottom of the conduction band into satellite valleys. Here, electrons show an increased effective mass with respect to low-lying states, leading to a reduction of carrier mobility and then to an enhanced THz transparency. A similar band-structure effect has been observed in Si and Ge giving rise to a comparable THz induced transparency \cite{sige}.

Recently, theoretical models have predicted a strong nonlinear THz response of 2-Dimensional (2D) metallic systems characterized by massless Dirac electrons. Their nonlinear response has been estimated to be higher than massive electron plasma in conventional metals \cite{conti,mikhailov,mikhailov2}. This strong nonlinearity can be qualitatively understood in terms of a simple calculation \cite{mikhailov}. Consider charge carriers having a Dirac dispersion $\epsilon (p)=V_{F}\cdot p=V_{F}\sqrt{p_{x}^2+p_{y}^2}$, where p is the linear momentum and $V_{F}$ is the Fermi velocity.
Under an oscillating THz Electric field $E(t)=E_0 \cos \nu t$, electrons gain (for zero scattering) a momentum $p(t)=-\frac{eE_0}{\nu} \sin \nu t$.  The band velocity (for instance along the $x$ direction), can be calculated through the equation $V_x=\partial \epsilon / \partial p_x$ and, for vanishing $p_y$,
$V_{x}(t)=-V_F sgn(\sin \nu t)$. Therefore, if $n_D$ is the Dirac surface density and neglecting the statistical distribution of carriers, this corresponds to an intraband Dirac current $J_{D}(t)=e n_D V_F sgn(\sin \nu t)$=$4/\pi$ e $n_D V_F$[$\sin \nu t$+1/3$\sin 3\nu t$+1/5$\sin 5\nu t$+....], which contains all odd harmonics $J_m$ (m=1,3,5..), with an amplitude decreasing as 1/m \cite{mikhailov2}. Although both the presence of scattering and the actual statistical distribution of electrons may modify the intensity ratio among harmonics \textcolor{black}{as theoretically calculated \cite{Naib, mikhailov3} and experimentally observed for instance in graphene \cite{Bowlan}}, the previous result still remains valid, and it is exclusively due to the peculiar charge-carrier Dirac energy/momentum dispersion.  

Graphene is the first material in which massless Dirac carriers have been predicted theoretically and soon experimentally observed \cite{graphene,graphene2}. Here, different experiments have further shown sizable nonlinear optical effects at THz frequency \cite{Bowlan, gnl,nelson,paul,drago}. 
However, Dirac electrons have been recently discovered in many other materials like 3D Dirac semimetals Cd$_3$As$_2$ and NaI$_3$ \cite{Cd3As2_cava, Na3I} giving rise to intensive investigations of their intriguing properties. Perhaps, one of the most important classes of Dirac systems is represented by 3D Topological Insulators. These materials are quantum systems characterized by an insulating electronic gap in the bulk, whose opening is due to strong spin-orbit interaction, and gapless surface states at their interfaces \cite{Moore}. Surface states in TIs are metallic, characterized by a Dirac dispersion, showing a chiral spin texture \cite{topo, topo2}, and protected from back-scattering by the time-reversal symmetry. Since their discovery, TI's have attracted a growing interest due to their potential application in quantum computing, \cite{Hasan,Kitaev}, THz detectors \cite{Zhang}, and spintronic devices \cite{Chen}. 
Linear THz spectroscopy has been applied to 3D TIs, in particular on thin films of HgTe and Bi$_2$Se$_3$, and there both Dirac single particle \cite{Shuvaev, Hancock, Armitage1, Armitage2} and collective (plasmon) \cite {NNano, AOM_Marta, ACS_Photonics} excitations $vs.$ temperature and applied magnetic field, have been observed. However, to our knowledge, THz nonlinear electrodynamics properties have been never investigated on 3D topological insulators so far.
In this paper we fill this gap, reporting on the first experimental observation of a strong nonlinear THz absorption in Bi$_2$Se$_3$ TI thin films.  Their electromagnetic response has been studied over seven decades of THz electric field amplitude (from 0.1 V/cm up to 1.5 MV/cm) by combining linear spectroscopy through conventional terahertz radiation with data achieved at the high-intensity SPARC$\_$LAB Linac-Based THz source in Italy. This source delivers broadband highly-intense THz pulses with femtosecond shaping \cite{Chiadroni,sparc2}.  In Bi$_2$Se$_3$ Topological Insulator we observed an induced electromagnetic transparency which increases and eventually saturates at high THz electric fields. 
This nonlinear THz behavior is associated with the presence of Dirac electrons onto the surfaces. 
Indeed, we do not observe any nonlinear effect in the same electric field interval in (Bi$_{0.9}$In$_{0.1}$)$_2$Se$_3$, a material having the same crystal structure of Bi$_2$Se$_3$ and showing instead a trivial topology characterized by a gas of massive (Schr\"odinger) electrons at the surface  \cite{Oh-PRL_In}.

\section{Results and Discussion}

Two thin films of Bi$_2$Se$_3$ were grown by Molecular Beam Epitaxy (MBE) on 0.5 mm thick sapphire (Al$_2$O$_3$) substrate. One film had a thickness t= 120 quintuple layers (QL), where 1 QL $\simeq$ 1 nm, the other one t= 60 QL \cite{ban1,ban2}. An additional film doped by In, (Bi$_{0.9}$In$_{0.1}$)$_2$Se$_3$, with t =  60 QL, which shows a trivial topology and a massive (Schr$\ddot{o}$dinger) electron gas with a similar surface density of Bi$_2$Se$_3$  \cite{Oh-PRL_In}, was grown on the same substrate for a sake of comparison. Each film was preliminarily characterized by transport and Hall measurements \cite{ban1,ban2}. Their linear response was further investigated by Fourier Transform terahertz spectroscopy (see also SI) \cite{NNano, AOM_Marta}.

In Fig.\ref{fig1} the experimental set-up is reported. Highly intense sub-picosecond THz pulses are produced at SPARC$\_$LAB as Coherent Transition Radiation (CTR) emitted by ultra-short ($\simeq$ 120 fs) high-brightness electron bunches \cite{Chiadroni}. THz radiation is reflected at 90$^\circ$ with respect to the electron beam direction and transmitted through a z-cut quartz window to an off-axis parabolic mirror. This mirror produces collimated radiation which is further reflected by a flat mirror at 45$^\circ$ and finally focalized on the film surface by a third off-axis parabolic mirror. 
A silicon beamsplitter mounted before the films, uses a portion of the beam to implement a differential detection system, allowing to remove shot by shot fluctuation effects of the source.
Finally, a pair of parallel wiregrid polarizers was used to tune the amplitude of  THz electric field over four decades: from 1 kV/cm to 1.5 MV/cm.  
We performed both integrated transmittance measurements (through a pyroelectric detector mounted behind the samples which collects the THz intensity transmitted by the film (substrate)), and spectrally resolved ones through a step-scan Michelson interferometer (see Fig.\ref{fig1}). Integrated and spectrally resolved transmittance at lower electric fields (from 0.1 V/cm to 1 V/cm) was measured by Fourier-transform spectroscopy with a conventional mercury lamp. 
As film transmittances have been normalized to the transmittances of the bare substrate, we have verified that the sapphire substrate response is independent of the amplitude of the THz field. Therefore, the nonlinear THz effects here observed are exclusively due to the Bi$_2$Se$_3$ Topological Insulator films.

Figure 2a, b shows the integrated transmittance at 300 K, normalized to that of substrate $vs.$ the THz electric field amplitude  E$_0$ for the 60 QL and 120 QL Bi$_2$Se$_3$ films. From 0.1 to 50 kV/cm the integrated transmittance (on the order of $60 \%$ for 60 QL and $50 \%$ for 120 QL) does not change appreciably. Therefore, this electric field range corresponds to the linear region in which the optical properties of Bi$_2$Se$_3$ are practically independent of the THz field.
For fields above this range the integrated transmittance follows a monotonous increasing behavior, indicating an enhancement of transparency. 
For fields above 1 MV/cm, instead it begins to saturate to a value of 70$\%$ (63$\%$) for 60 QL (120 QL) film, corresponding to an enhanced transparency of about 20$\%$ with respect to the value in the linear region.

In the insets of Fig. 2a,b we show the frequency resolved transmittance for both samples at three different electric field amplitudes: 0.1 V/cm (as measured by a Michelson interferometer coupled to a Mercury source), 0.5 MV/cm and 1 MV/cm (these last values through the THz source at SPARC\_LAB). 
At the lowest field both films show a transmittance which slightly decreases for frequency $\nu \to 0$. This behavior is the signature of a free electron (Drude) absorption, which is associated mainly with Dirac surface states \cite{Armitage1, Armitage2}, although a contribution from a 2D massive electron gas due to band-bending effects, cannot be ruled out, in particular at room-T \cite{Stauber}. Moreover, in the high-frequency part of the transmittance spectrum (around 1.8 THz) one observes a minimum which is related to the presence of the bulk $\alpha$ phonon absorption superimposed to the Drude term (see SI for details). Spectrally resolved transmittances are quantitatively similar to the ones measured on films belonging to the same batch in Ref. \onlinecite {Armitage1, Armitage2, NNano, AOM_Marta}.
By increasing the THz field, transmittances become more flat and, at 1 MV/cm, enhance their value to about $80$\%$ ($75$\%$) for the 60 (120) QL films.

The integrated transmittance as a function of the THz electric field amplitude E$_{0}$ can be described through a phenomenological saturable absorption model:

\begin{equation}
T(E_0)=T_{ns}\frac{ln[1+T_{lin}/T_{ns}(e^{E_{0}^2/E_{sat}^2}-1)]}{E_{0}^2/E_{sat}^2}
\label{equation}
\end{equation}

\noindent where $T_{lin}$ and $T_{ns}$ are the linear and the non-saturable integrated transmittances, and $E_{sat}$ is the THz electric field saturation value. If one fixes $T_{lin}$ and $T_{ns}$ from data ($T_{lin}= 0.59 \ \%$ and $T_{ns}=0.69 \ \%$ for 60 QL film, $T_{lin}= 0.51 \ \%$ and $T_{ns}=0.65 \ \%$ for 120 QL film), one obtains from the fit (blue point-dashed curves in Fig.\ref{fig2}a, b), comparable values of $E_{sat}$ for both samples: 0.32 MV/cm for 120 QL and 0.31 MV/cm for 60 QL, which correspond to a saturation fluence $F$ of about 50 $\mu$J/cm$^2$.

The constant value of $E_{sat}$ as measured on different thickness (60 and 120 QL) Bi$_2$Se$_3$ films, is a clear signature that the nonlinear absorption in topological insulators is a surface property and does not depend on their bulk characteristics.
Moreover, such a saturation fluence is comparable to that measured in doped graphene ($F$=20 $\mu$J/cm$^2$) \cite{nelson}.\\

In order to properly assign the nonlinear electromagnetic response observed in Bi$_2$Se$_3$ to Dirac electrons, we have investigated also the optical response of a (Bi$_{1-x}$In$_{x}$)$_2$Se$_3$ film with x=0.1. While Indium substitution does not change the crystal structure, it induces a Quantum Phase Transition from a topological to a trivial phase  for an In content larger than 0.045  \cite{Armitage2}. Moreover, the (Bi$_{0.9}$In$_{0.1}$)$_2$Se$_3$ film shows, due to band-bending effects, a gas of massive electrons having a surface density ($n_{M}$=2.5 $\times$ 10$^{13}$ cm$^{-2}$), comparable to that of 2D Dirac carriers as observed in Bi$_2$Se$_3$ \cite{Oh-PRL_In}. 
At variance with Bi$_2$Se$_3$, the integrated transmittance of (Bi$_{0.9}$In$_{0.1}$)$_2$Se$_3$, which is shown in Fig.\ref{fig2}c, is completely flat over seven decades of the THz electric field amplitude. \textcolor{black}{Moreover, the spectrally resolved transmittances both at 0.1 V/cm and 1 MV/cm are superimposed within our experimental sensitivity.}
These results undoubtedly indicate that the strong nonlinear electromagnetic response previously observed in Bi$_2$Se$_3$ must be attributed to the 2D gas of Dirac electrons present at the surface of Topological Insulators.

In graphene, the electromagnetic induced transparency has been interpreted in terms of a combination of two mechanisms: Harmonic generation and a strong decrease of carrier mobility ($i.e.$ an increase of carrier scattering rate), due to the opening of new scattering channels for the accelerated carriers \cite{nelson, paul, Bowlan}. 

\textcolor{black}{In order to clarify the transparency microscopic mechanism in Bi$_2$Se$_3$ topological insulator, we have performed a further experiment on a 60 QL thick film with the aim of looking for a specific harmonic signal. 
In particular, a pulse centered at 1 THz with a maximum field of about 300 kV/cm (selected from the broad SPARC spectrum by a bandpass THz filter, see Materials and Method section), has been used to illuminate the 60 QL Bi$_2$Se$_3$ film. Let us observe that at 1 THz the SPARC source has its maximum intensity (see SI). The transmitted intensity was collected through a bandpass filter centered at 3 THz (for the optical scheme see Fig. 3a) and finally measured through a pyroelectric detector. 
By varying the incident intensity ($i.e.$ the incident electric field) through a couple of THz polarizers we were able to measure the transmitted intensity from an incident field E$_0$ of about 1 kV/cm to a maximum field of 300 kV/cm. This optical configuration may capture the third harmonic generation (1 THz $\rightarrow$ 3 THz) in Bi$_2$Se$_3$ films.
The transmitted intensity (normalized to its maximum value in Fig. 3b), first increases smoothly at the low-field values. Around 50 kV/cm the intensity rapidly grows, scaling with (E$_0$/E$_{max}$)$^6$ (black dashed line in Fig. 3b), where E$_{max}$ =300 kV/cm. The dependence on the sixth power of E$_0$ suggests (see for instance Ref. \onlinecite{Matsunaga}), a third harmonic generation process. 
Let us note that 50 kV/cm corresponds to the field where the electromagnetic momentum gained by electrons is comparable with the Fermi momentum. The efficiency of third harmonic generation process has been measured through the ratio $\epsilon=I(3\nu)/I(\nu)$, where I(3$\nu$) (I($\nu$)), is the transmitted intensity of a sample at the frequency 3$\nu$ ($\nu$).This ratio at $\nu$= 1 THz  is about 1$\%$ in quite good agreement with graphene  \cite{Bowlan}. $\epsilon$ depends on four parameters of the Dirac electron gas: the Fermi velocity v$_F$, the Fermi energy E$_F$, the Dirac charge density n$_D$ and the actual scattering rate of charge-carriers $\Gamma$. $\epsilon$ has been calculated theoretically for graphene in Ref. \onlinecite{mikhailov3} and theory can be straightforward extended to Bi$_2$Se$_3$.
Indeed, the first three parameters are well known from transport and photoemission experiments performed on films of the same batch \cite{ban1, Armitage2}: $V_F$=5 $\pm$ 1 $\times$ $10^{7}$ cm/s, $E_F$=380 $\pm$ 10 meV, and $n_{D}=3\pm1\times10^{13}$ cm$^{-2}$. 
Therefore, by using Eq. 11 in Ref. \onlinecite{mikhailov3} a strong reduction of the efficiency to $\sim$ 6 $\%$ from the ideal value of 1/3 is already obtained at 300 kV/cm and 1 THz, if the scattering rate is set to its linear value of 3.1 THz. An efficiency of $\sim$ 1 $\%$ is finally achieved by considering a further field renormalization of the scattering rate. Indeed, $\epsilon$ is a rapid decreasing function of $\Gamma$ (inset in Fig. 3b), and takes a value around 1 $\%$ when $\Gamma$ is $\sim$ 5.5 THz.  
This suggests a quite strong enhancement of the scattering rate of Dirac electrons $vs.$ E$_0$ as already observed for graphene in Ref. \onlinecite{nelson, Naib, mikhailov3}. 
In graphene, several mechanisms contribute to the scattering of Dirac electrons: Electron-electron scattering, short and long range impurity scattering, and optical-phonon interaction \cite{nelson, winnerl}. The relative importance of these mechanisms depends on the Fermi energy and temperature. As a matter of fact, to a strong THz pulse corresponds an overall increase of the scattering rate and then a reduction of harmonic conversion efficiency. Although, in Bi$_2$Se$_3$ poor information is present about the relative importance of different scattering mechanisms and on their characteristic temporal scale, one can expect that a similar renormalization of the scattering rate takes place even at higher fields. However, by taking into account the quite large dependence of $\Gamma$ with T \cite{AOM_Marta}, one can foresee an increase of the efficiency at low-temperature that could be further improved through a proper control of film growing process to reduce the impurity and defect scattering contributions.}

\section{Conclusions}

The electromagnetic response of Bi$_2$Se$_3$ thin films has been investigated over seven decades of THz electric field: From 0.1 V/cm by means of conventional Fourier Transform terahertz spectroscopy to 1.5 MV/cm by using the linac-based SPARC\_LAB terahertz source. 
We observed for the first time a a strong reduction of the absorption of Bi$_2$Se$_3$ Topological Insulator for an increasing THz field which determines an electromagnetic induced transparency in this material. 

The induced transparency is determined only by the surface states of Bi$_2$Se$_3$ as films with different thickness shows exactly the same THz behavior. Moreover, a similar experiment performed on a (Bi$_{0.9}$In$_{0.1}$)$_2$Se$_3$ film, which presents a trivial topological phase characterized by a gas of massive electrons, shows an absorption which does not depend on the THz field amplitude. This demonstrates that strong nonlinear effects in Bi$_2$Se$_3$ are driven by massless Dirac electrons at the surface. 

\textcolor{black}{We directly observe at 300 K a harmonic generation process as demonstrated by the harmonic signal at 3 THz when a Bi$_2$Se$_3$ film is nonlinearly stimulated by a strong pulse at 1 THz. The efficiency of harmonic generation is on the order of 1 $\%$ at variance with the nominal value of 1/3. This strong renormalization is mainly determined by the charge-carriers scattering rate so working at low-temperature and reducing the scattering through improved growth conditions, may increase the harmonic conversion.} 

In conclusion, the possibility to control light by light in the THz regime is an actual subject of intense study to implement compelling applications in THz technology, like ultrafast THz tabletop sources, quantum cascade lasers and ultrafast THz communications based on optical bistability. In this regard, the strong nonlinear THz properties observed in Bi$_2$Se$_3$ Dirac material, could open promising perspectives in the tumultuous field of terahertz technologies.   

\section{Materials and Methods}
\subsection{Topological Insulator films} 
The high-quality (Bi$_{1-x}$In$_{x}$)$_2$Se$_3$ thin films were prepared by molecular beam epitaxy using the standard two-step growth method developed at Rutgers University \cite{ban1,ban2}. The 10$\times$10 mm$^2$ Al$_2$O$_3$ substrates were first cleaned by heating to 750 $^\circ$C in an oxygen environment to remove organic surface contamination. An initial three quintuple layers of Bi2Se3 were deposited on the substrates at 110 $^\circ$C,  which was then followed by heating to 220 $^\circ$C helping further  to achieve the target thickness. The crystallization of the films during the growth was monitored by reflection high energy electron diffraction (RHEED). The Se:(Bi/In) optimal flux ratio was 10:1 for the deposition. A pre-control Bi/In flux ratio was performed to achieve the desired In concentration. Once the films were cooled, they were removed from the vacuum chamber, and vacuum-sealed in plastic bags within two minutes, then shipped to the University of Rome.\\

\subsection{High field THz generation and measurements}
The ultra-relativistic electron bunches 650 pC charged with a time duration of $\simeq$120 fs and 10 Hz of repetition rate was used to generate nearly single-cycle THz pulses as coherent transition radiation at SPARC\_LAB \cite{Chiadroni,sparc2}. 
The THz electric field in the focal point was estimated in two ways. The first estimate consists in using the nominal sensitivity of the pyroelectric detector (140 kV/W) which has been experimentally tested at 970 GHz through a Schottky Diode (Virginia Diode) emitting a power of 1 mW. In the second estimate we calculated the THz electric field in the focal point produced by the CTR SPARC source by the THz-Transport simulation program \cite{thz_transport}. In this calculation we take into account both the actual (finite) size of CTR target, the actual transmittance of the z-cut quartz window and the optical properties of the propagation optics. Both methods provide comparable THz electric fields in the focal point having a highest THz field amplitude of 1.5  MV/cm.\\ 

The THz radiation coming from the CTR source was further separated in two beams by a high-resistivity Si beamsplitter at 45$^{\circ}$(see Fig.1). The transmitted beam propagates towards the films, while that reflected one towards a pyroelectric detector. The THz integrated spectra were measured placing just behind the films a pyroelectric detector, meanwhile the THz spectra have been measured by a step-scan Michelson Interferometer, having a 24 $\mu m$ Mylar pellicle beamsplitter. All radiation channels were equipped with THz-I-BNC GENTEC-EO pyroelectric detectors. 

We used a differential detection technique to reduce the shot-to-shot fluctuations of the SPARC THz source. In particular, both the integrated signal and the spectrally resolved one were normalized to the signal as measured by the reference pyroelectric detector. This technique provides an error-bars on the transmittance (both integrated and spectrally resolved) on the order of 0.5 $\%$. 

The THz filters used in the harmonic detection experiment have been acquired from Tydex (http://www.tydexoptics.com) and their optical response controlled by a Michelson interferometer. The THz radiation transmitted by the series 1 THz filter, 60 QL film and 3 THz filter has been finally collected by a THz-I-BNC pyroelectric by GENTEC-EO.\\ 

\section{Acknowledgements}
We thank S.A. Mikhailov for useful discussion about harmonic generation theory. M. B., N. K. and S. O. thank for the financial support the Office of Naval Research (N000141210456) and the Gordon and Betty Moore Foundation’s EPiQS Initiative through Grant GBMF4418.

\section{Author Contributions} 
M. B., N. K. and S. O. fabricated and characterized the (Bi$_{1-x}$In$_{x}$)$_2$Se$_3$ films.  F.G., E.C., A.R., M.C.G., and S.L. carried out the terahertz experiments and data analysis. M.B., E. C., M.C., D.D., G.D., M.F., R.P., C.V., F.V., A.C., A.M., M.P. managed the SPARC machine during the THz experiments.
S. L. planned and managed the project with inputs from all the co-authors. F.G. A. P. and S.L. wrote the manuscript. All authors extensively discussed the results.

\section{Additional Information}
The authors declare no competing financial interests. 
Correspondence and requests for materials should be adressed to S.L. (stefano.lupi@roma1.infn.it).

\begin{widetext}

\begin{figure}[h]   
\begin{center}    
\includegraphics[width=15cm]{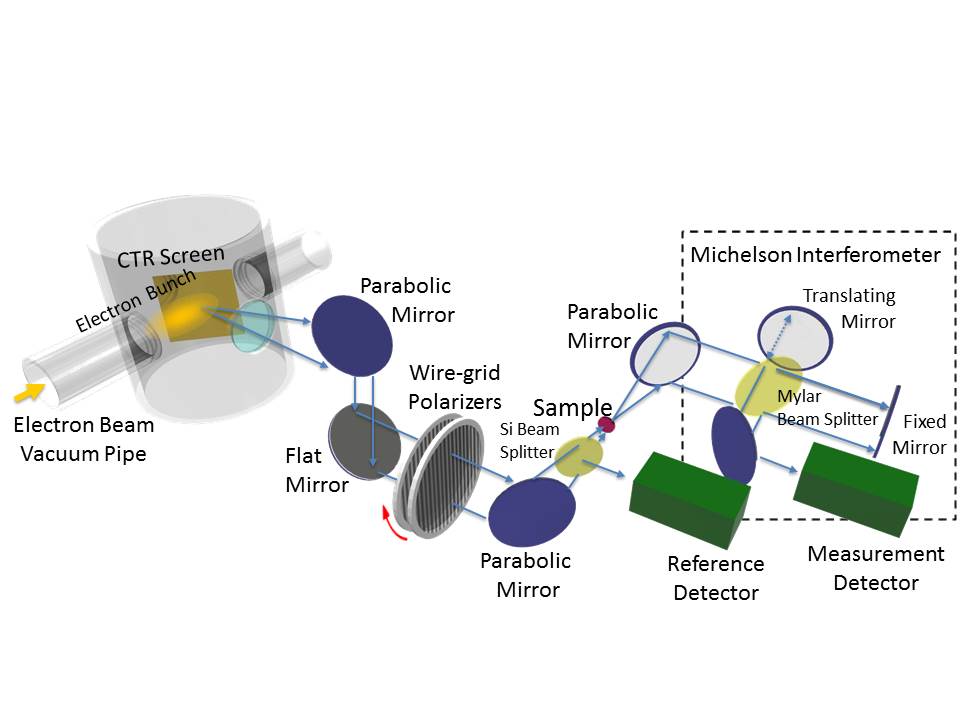}
\caption{ {\bf Scheme of experimental setup at SPARC THz source.}
The ultra-short electronic bunches coming from SPARC LAB photoinjector interacting with a metallic screen produce highly intese sub-picosecond Coherent-Transition-Radiation THz pulses. THz radiation (blue arrows), emitted at 90$^\circ$ with respect the electron propagation direction, is transmitted by a z-cut Quartz window and collected and collimated by means of an axis-off parabolic mirror. A further flat mirror was used to reflect the THz radiation up to the optical table where a second off-axis parabolic mirror focalized the THz pulses on film samples. A pair of parallel wiregrid polarizers (QMC Inc.) have been used to tune the amplitude of the THz electric field over four decades: from 1 kV/cm to 1.5 MV/cm.   A further, twin, off-axis parabolic mirror is finally used for illuminating a Michelson interferometer equipped with a GENTEC-EO pyroelectric detector which has been used for measuring the spectrally resolved transmittance. A further GENTEC-EO pyroelectric detector was mounted before the films to implement a differential detection (see SI), in order to remove shot by shot fluctuation effects of the SPARC THz source.
Integrated transmittances were measured substituting the Michelson interferometer with another GENTEC-EO pyroelectric detector which is mounted just behind the films.}
\label{fig1} 
\end{center}
\end{figure}

\begin{figure}[h]
\centering
\includegraphics[scale=0.7]{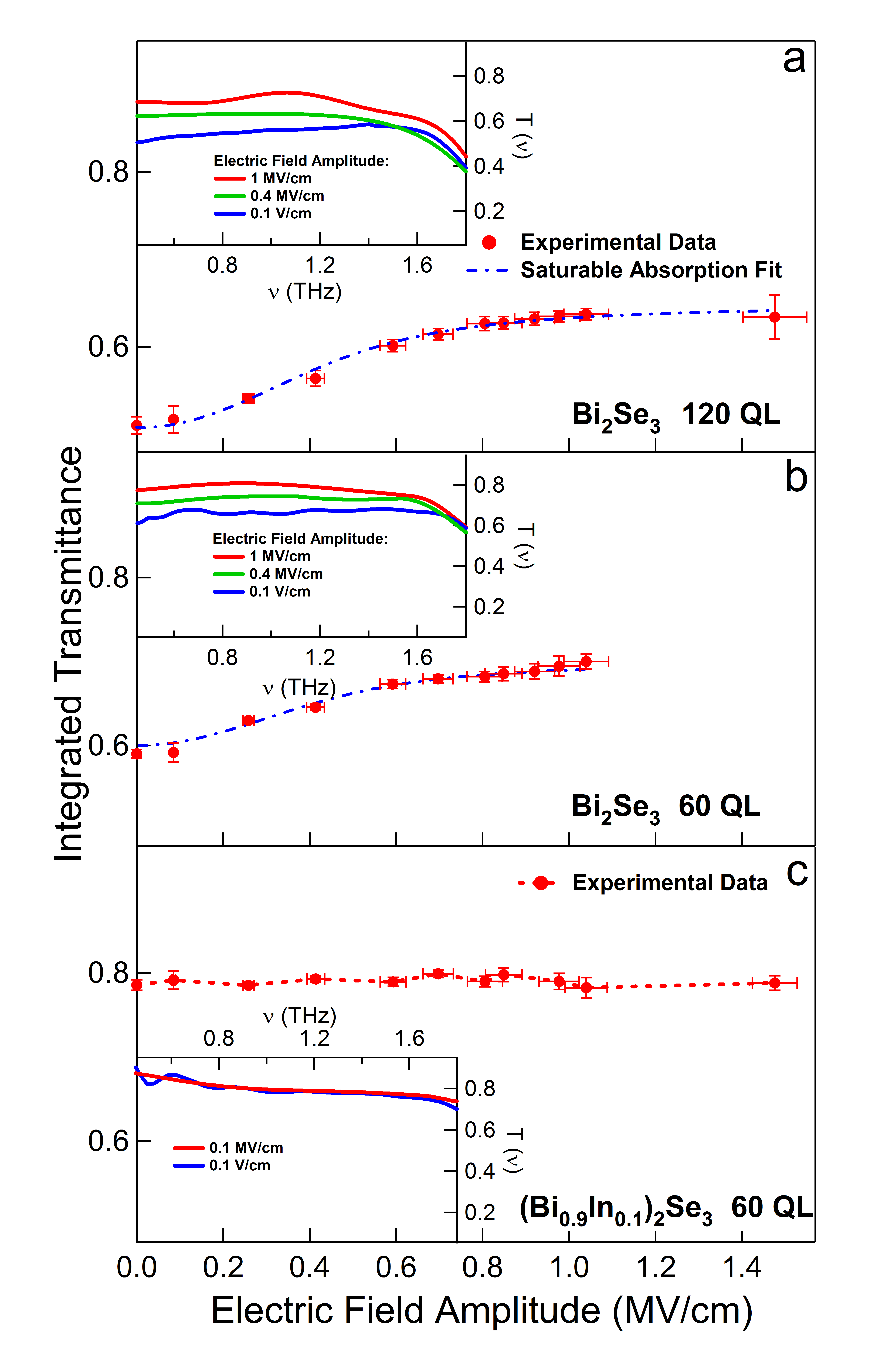}  
\caption{  {\bf THz nonlinear behaviors of the Bi$_2$Se$_3$ Topological Insulator.}
Integrated transmittance of Bi$_2$Se$_3$ 120 QL (\textbf{a}), Bi$_2$Se$_3$ 60 QL (\textbf{b}) and (Bi$_{0.9}$In$_{0.1}$)$_2$Se$_3$ 60 QL (\textbf{c}) films, respectively, as a function of the incident THz electric field amplitude E$_0$. Experimental data, and a saturable absorption fit (see text) are represented by red-dots, dashed-dotted-blue line, respectively. 
Insets: spectrally resolved transmittance curves measured (solid lines) at 1 MV/cm (red curve), 0.4 MV/cm (green curve), and 0.1 V/cm (blue curve)  for Bi$_2$Se$_3$ 120 QL (\textbf{a}) and Bi$_2$Se$_3$ 60 QL (\textbf{b}).The spectrally resolved transmittance of (Bi$_{0.9}$In$_{0.1}$)$_2$Se$_3$ 60 QL at 0.1 V/cm (blue curve), and 1 MV/cm (red curve) are shown in the inset of Fig. 2(\textbf{c}).\textcolor{black}{They are superimposed in the limit of our sensitivity. The slow modulation in the spectrally-resolved transmittances are related to a non-perfect compensation of water absorption in the THz range.}} 
\label{fig2}
\end{figure}

\begin{figure}[h]
\centering
\includegraphics[scale=0.8]{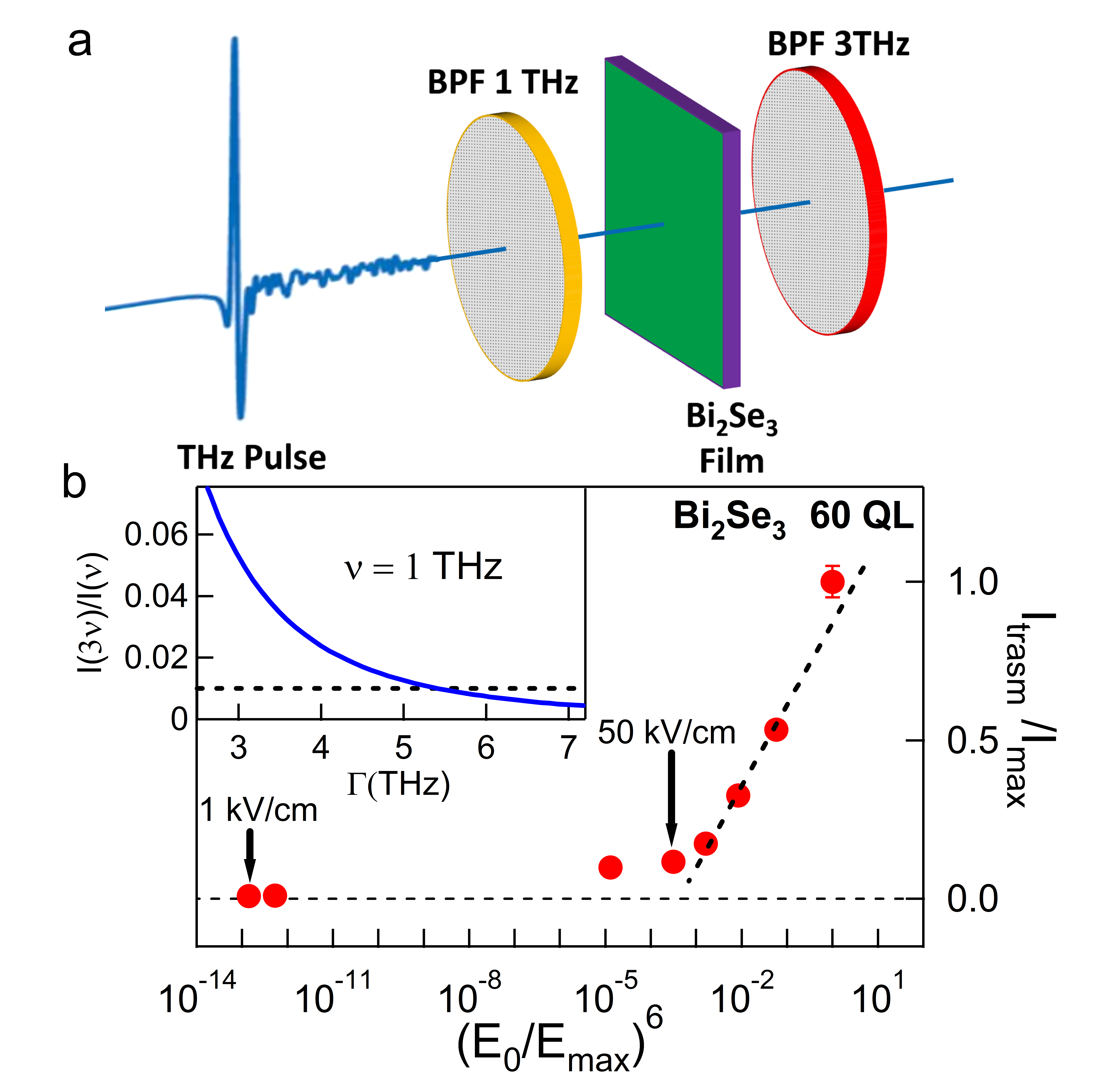}  
\caption{  {\bf Third harmonic generation in Bi$_2$Se$_3$ Topological Insulator.} Optical scheme for a third harmonic measurement. An bandpass optical filter selects from the broad SPARC THz spectrum a pulse centered at 1 THz and having a relative bandwidth of nearly 15$\%$. This pulse, with a maximum field of 300 kV/cm, illuminates a 60 QL Bi$_2$Se$_3$ film. The transmitted intensity is collected through a filter centered at 3 THz and having the same relative bandwidth and finally measured through a pyroelectric detector (\textbf{a}). Above nearly 50 kV/cm the transmitted intensity (normalized to its maximum value) follows a (E$_0$/E$_{max}$)$^6$ dependence (where E$_{max}$=300 kV/cm), suggestings a third harmonic conversion process (\textbf{b}). In the inset of panel b, the efficiency $\epsilon$ of third harmonic generation is represented $vs.$ the scattering rate $\Gamma$. The measured efficiency of 1 $\%$ is obtained for $\Gamma$ $\sim$ 5.5 THz.} 
\label{fig3}
\end{figure}

\end{widetext}

\end{document}